\documentclass{ar-1col}
\usepackage[numbers,sort&compress]{natbib}
\usepackage{amssymb}
\usepackage{amsmath}
\usepackage{geometry}
\let\oldmarginpar\marginpar
\renewcommand\marginpar[1]{\-\oldmarginpar[\raggedleft\footnotesize #1]%
{\raggedright\footnotesize #1}}

\newcommand{\bea}{\begin{eqnarray}}
\newcommand{\eea}{\end{eqnarray}}

\renewcommand{\epsilon}{\varepsilon}

\newcommand{\sign}{\mathrm{sign}}
\def\avg#1{\left\langle#1\right\rangle}
\def\Fig#1{Fig.~\ref{#1}}
\def\Eq#1{Eq.~(\ref{#1})}

\def\Tr{\mathrm{Tr}}

\newcommand{\tabincell}[2]{
\begin{tabular}{@{}#1@{}}#2\end{tabular}
}

\jname{Annu. Rev. Condens. Matter Phys.}
\jyear{2019}

\begin{document}

\title{Sign-Problem-Free Fermionic Quantum Monte Carlo: Developments and Applications}

\author{Zi-Xiang Li$^{1,2}$ and Hong Yao$^{1,3}$
\affil{$^1$Institute for Advanced Study, Tsinghua University, Beijing 100084, China;\\
email: yaohong@tsinghua.edu.cn}
\affil{$^2$Department of Physics, University of California, Berkeley, CA 94720, USA}
\affil{$^3$State Key Laboratory of Low Dimensional Quantum Physics, Tsinghua
University, Beijing 100084, China}}

\begin{abstract}
Reliable simulations of correlated quantum systems, including high-temperature superconductors and frustrated magnets, are increasingly desired nowadays to further understanding of essential features in such systems. Quantum Monte Carlo (QMC) is a unique numerically-exact and intrinsically-unbiased method to simulate interacting quantum many-body systems. More importantly, when QMC simulations are free from the notorious fermion-sign problem, they can reliably simulate interacting quantum models with large system size and low temperature to reveal low-energy physics such as spontaneously-broken symmetries and universal quantum critical behaviors. Here, we concisely review recent progresses made in developing new sign-problem-free QMC algorithms, including those employing Majorana representation and those utilizing hot-spot physics. We also discuss applications of these novel sign-problem-free QMC algorithms in simulations of various interesting quantum many-body models. Finally, we discuss possible future directions of designing sign-problem-free QMC methods.
\end{abstract}

\maketitle

\tableofcontents

\section{Introduction to quantum Monte Carlo and fermion-sign problem}

Understanding physical properties of interacting quantum many-body systems with strong correlations is of central importance in modern condensed matter physics \cite{Wenbook,Fradkinbook,Sachdevbook} as well as in related fields such as nuclear physics \cite{nuclear1} and quantum chromodynamics \cite{QCD1,QCD2}. However, mainly due to the exponentially large Hilbert space of a quantum many-body system and the lack of small expansion parameters, most strongly-interacting quantum many-body systems in two or higher spatial dimensions are beyond the solvability of known controlled analytical methods. Unbiased and efficient numerical methods are highly demanded to investigate low-energy and long-distance physics of interacting quantum many-body systems, especially those exhibiting intriguing physical properties such as high-temperature superconductivity and fractionalized excitations. Among various numerical approaches available nowadays, quantum Monte Carlo (QMC) has been established as one of the most important approaches to simulate many-body systems in two and higher dimensions \cite{QMC1,QMC2,QMC3,QMC4,QMC5,QMC6,QMC7,QMC8,QMC9,QMC10,QMC11}.
One unique feature of QMC is that it does not rely on variational principle and is then intrinsically unbiased.
Specifically, QMC simulations employ stochastic sampling, e.g. using Metropolis algorithm, in small but representative subset of Hilbert space, instead of performing direct computations in the whole Hilbert space. One major goal of simulating quantum many-body systems by QMC is to obtain unbiased and accurate results within computational time that scales only {\it polynomially} with the system's particle number (despite that the Hilbert space being {\it exponentially} large) so that reliable study of systems with large size and low temperature is plausible.

\begin{figure}[t]
\includegraphics[height=1.2in]{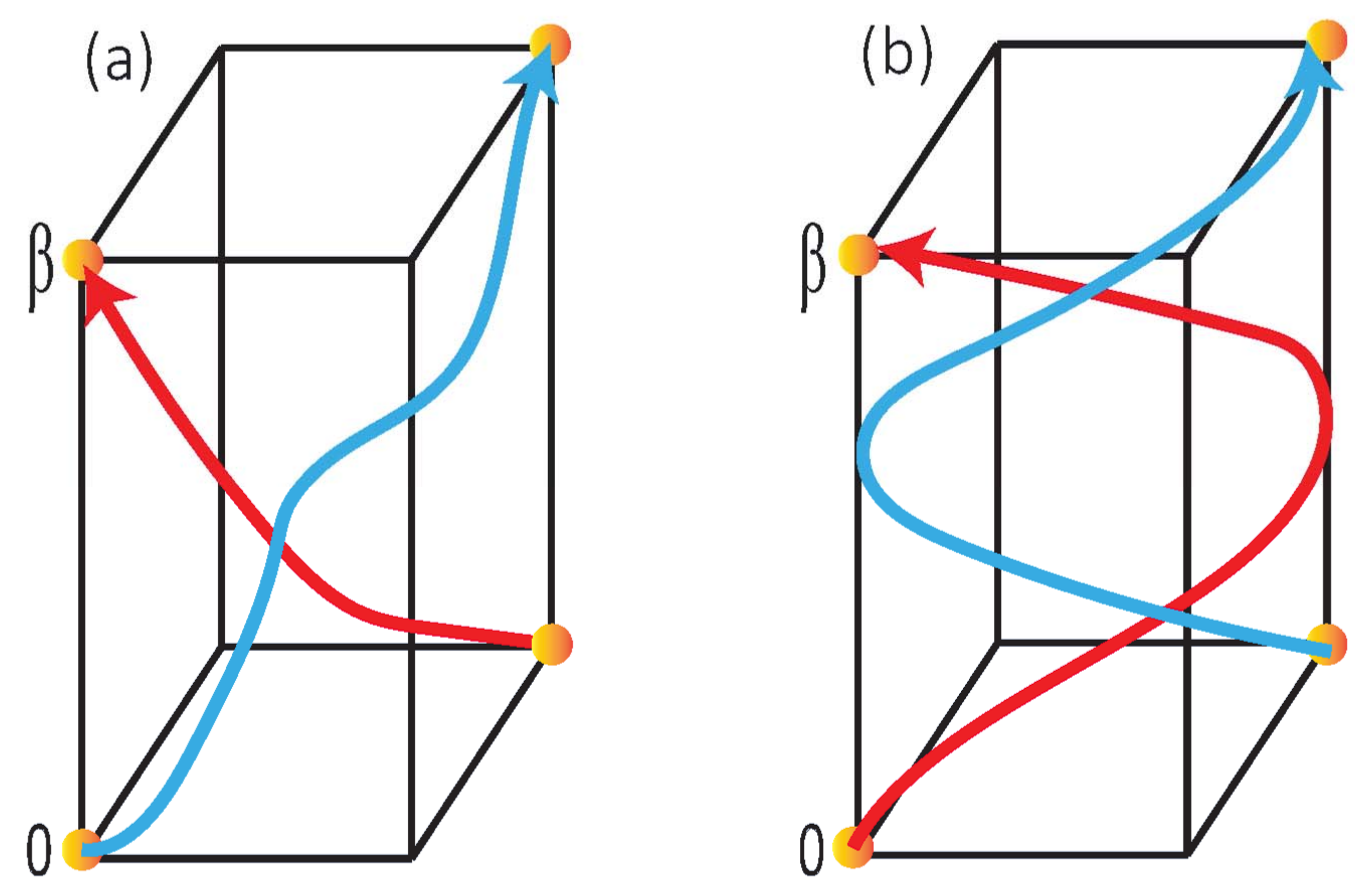}
\caption{ Schematic world-line configurations in a (2+1)-dimensional space-time lattice. (a) The weight of the world-line configuration is negative when fermions exchange for odd times; (b) The weight of the world-line configuration is positive when fermions exchange for even times. }
\label{signproblem}
\end{figure}

The Monte Carlo method was originally proposed to study classical systems, the basic idea of which is stochastic sampling of the classical configurations whose stochastic probabilities are the Boltzmann weights in the partition function \cite{Metropolis-1949}. Classical Monte Carlo has achieved great successes in understanding properties of classical many-body systems. As a quantum system in $d$ spatial dimensions can usually be mapped to a classical one in $d$+1 dimensions by employing the path integral representation, Monte Carlo methods can be formally applied to simulate quantum many-body systems. Specifically, the partition function of a quantum many-body system with Hamiltonian operator $\hat H$ and inverse temperature $\beta\equiv 1/T$ is given by:
\bea
Z=\Tr\left[e^{-\beta \hat H}\right],
\eea
where the trace is over the system's whole Hilbert space which grows exponentially with system size. Directly obtaining the quantum partition function for an interacting many-body system is usually neither feasible nor necessary. One key step in QMC simulations is to map such quantum partition function into a summation of classical Boltzmann weights $Z=\sum_c w(c)$; which $c$ labels classical configurations defined in a certain way. For instance, utilizing the spirit of path integral, the quantum partition function can be expressed as a sum over all possible world-line configurations $c$ with statistical weight $w(c)$. This observation is the foundation of a typical QMC algorithm, namely world-line QMC \cite{worldline1,worldline2}. In the world-line QMC, $c$ represents the configuration of all particles' world lines in space and imaginary-time, as schematically shown in \Fig{signproblem}. The putative Monte Carlo procedure is to perform stochastic sampling of the world-line configurations according to the ``probability'' $w(c)$. Indeed, in classical Monte Carlo, the Boltzmann weights are always positive so that stochastic sampling can be carried out directly by interpreting Boltzmann weights as probabilities. However, in quantum Monte Carlo simulations of fermionic systems as well as frustrated bosonic systems, Boltzmann weights $w(c)$ are not necessarily positive-definite due to the fermionic nature of constituting particles \cite{sign1}. For example, a world-line configuration which has an odd number of fermion exchanges would lead to a negative Boltzmann weight, as shown in \Fig{signproblem}(a). The appearance of non-positive Boltzmann weights lies in the intrinsic difference between classical and quantum systems. Because negative Boltzmann weight cannot be treated directly as probability, the minus sign issue severely hampers the application of QMC to reliably study interacting quantum systems with a large number of particles, which we shall illustrate below.

One straightforward way of dealing with the minus sign in Boltzmann weights is to employ the absolute value of Boltzmann weight $|w(c)|$ as probability of sampling. The expectation value of an observable represented by the operator $\hat O$ can be computed as follows:
\bea\label{signaverage}
 \langle \hat O \rangle = \frac{\sum_c w(c) O(c)}{\sum_c w(c)} = \frac{\sum_c O(c)\sign(c)|w(c)|/\sum_c |w(c)|}{\sum_c \sign(c)|w(c)|/\sum_c |w(c)|} = \frac{\langle \hat O \rangle_{|w|}}{\avg{\sign}_{|w|}},
\eea
where $w(c)=\sign(c)|w(c)|$; namely $\sign(c)$ labels the sign of $w(c)$. This procedure allows QMC can be implemented even when the Boltzmann weights in question are not positive-definite. However, the configurations with negative sign and positive sign nearly cancel with each other such that $\avg{\sign}_{|w|}$ is exponentially small with the system size; this results in exponentially large statistical errors in evaluating the observable $\langle \hat O \rangle$. To be more explicit, we illustrate this using the world-line representation. In the world-line algorithm, the sign average $\avg{\sign}_{|w|}$ decays exponentially with the system's particle number $N$ and inverse temperature $\beta$: $\avg{\sign}_{|w|} = \exp(-N \beta \Delta f)$, where $\Delta f>0$ represents the free energy density difference between the fermionic quantum system under consideration and the corresponding bosonic one \cite{Troyer-2005}. Consequently, the statistical errors generated in evaluating $\langle \hat O \rangle$ shall grow exponentially with the particle number $N$ and inverse temperature $\beta$, and is proportional to $\frac1{\sqrt{M}} \exp(N \beta \Delta f)$, where $M$ represents the number of sampling steps in Monte Carlo simulations \cite{Troyer-2005}. To achieve a given accuracy in evaluating $\langle \hat O \rangle$, the required number of Monte Carlo sampling steps or equivalently the computational time grows exponentially with $N\beta$, the product of the system size with the span of imaginary time. Even equipped with the state-of-the-art supercomputers, this exponential growth of required computational time severely prevents the application of QMC to reliably study fermionic systems with large systems size and low temperature. This is the so-called {\it fermion-sign problem}. Even though whether the fermion-sign problem occurs depends on representation and algorithm, it has been proven that the general solution of the fermion-sign problem has been proven to be nondeterministic polynomial (NP)-hard \cite{Troyer-2005}. This implies that achieving a generic solution of fermion-sign problem is believed to be extremely difficult, if not impossible. Consequently, fermion-sign problem is a central obstacle in applying QMC simulations to study quantum many-body systems of fermions.

Fortunately, it is still {\it possible} that the fermion-sign problem can be circumvented in specific quantum models by designing appropriate QMC algorithms. The fermion-sign problem generically appears in QMC simulations of fermionic systems by world-line algorithm. To avoid the sign problem and simulate quantum models more efficiently, various fermionic QMC methods other than the world-line one for quantum lattice models \cite{QMC3,QMC6,QMC7,QMC8,fixednode1,fixednode2,constraintpath1,constraintpath2} and for impurity models \cite{QMC9,impurity1,CTQMCimpurity1,CTQMCimpurity2} have been proposed. One may roughly classify most of these approaches into two distinct types. The algorithms of the first type are based on Suzuki-Trotter decomposition of the imaginary time \cite{QMC1,QMC3,QMC6,QMC7,QMC8}. The imaginary time is discretized such that a $d$-dimensional lattice is transformed to a ($d$+1)-dimensional space-time lattice. A typical fermionic QMC algorithm in this class is the determinant QMC (DQMC). Instead of discretizing imaginary time, algorithms of the second type map the quantum partition function to a classical summation by performing diagramatic expansions \cite{CTQMC1,CTQMC2,CTQMC3,CTQMC4,CTQMC5,CTQMC6,CTQMC7,CTQMC8}.

Here we shall mainly focus on the DQMC. In the DQMC algorithm, fermion-sign problems may be solved for certain quantum models because one can sum the weights of a subset of fermion-exchange processes into a matrix determinant which could be positive-definite. The first DQMC approach of simulating interacting fermion models was proposed in 1981 by Blankenbecler, Scalapino, and Sugar (BSS) \cite{QMC3}. The basic idea of the BSS algorithm is to convert an interacting fermion model into a problem of free fermions coupled with classical auxiliary fields, and then sample over auxiliary-field configurations with the probabilities that is a auxiliary-field-configuration-dependent determinant obtained by integrating out free fermions. Specifically, to perform DQMC simulations of an interacting quantum model described by the Hamiltonian $\hat H$, one employs the Suzuki-Trotter decomposition as follows:
\bea
Z=\Tr \left[\prod_{n=1}^{N_\tau} \exp(-\Delta\tau \hat{H})\right]
= \Tr \left[\prod_{n=1}^{N_\tau} \exp(-\Delta\tau \hat{H}_0) \exp(-\Delta\tau \hat{H}_I)\right]+\mathcal{O}(\Delta\tau^2),
\eea
where $\Delta \tau  = \beta/N_{\tau}$ and $\hat H=\hat H_0+\hat H_I$ with $\hat H_0$ representing the non-interacting part and $\hat H_I$ the four-fermion interactions. Typically, the non-interacting part can be written as $\hat H_0=c^\dag H_0 c=c^\dag_i H_0(i,j) c_j$, where $c^\dag_i$ are fermion creation operators with $i$ labeling a collection of relevant indices such as site, spin, and orbital; and $H_0$ is a matrix. The errors caused by the Suruki-Trotter decomposition is proportional to $\Delta\tau^2$, and can therefore be made sufficiently small by decreasing the time slice $\Delta\tau$. The standard Hubbard-Stratonovich (HS) decomposition, which is based on the Gaussian integral, can be applied to decouple four-fermions interactions $\hat H_I$: $\exp(-\Delta\tau \hat H_I) = \sum_{\phi_n} \exp[-\Delta\tau\hat{h}(\phi_n)]$, where $\hat{h}(\phi_n)=c^\dag h(\phi_n) c$ represent quadratic fermion operators with the matrix $h(\phi_n)$ depending on the configurations of auxiliary fields $\phi_n$ at imaginary time $\tau_n=n\Delta \tau$. Note that the spatial (such as site and bond) dependence of $\phi_n$ is implicit here. Finally, tracing out fermions can be done because only quadratic fermion operators appear in the exponentials, which gives rise to a determinant which depends on the auxiliary-field configurations:
\bea{\label{determinant}}
Z = \sum_{\{\phi_n\}} \Tr\left[ \prod_{n=1}^{N_\tau} e^{-\Delta\tau \hat{H}(n)}\right]
= \sum_{\{\phi_n\}}\det\left[\mathbb{I}+ e^{-\Delta\tau H(n)}\right],
\eea
where $\hat H(n)=\hat H_0+\hat h(\phi_n)$ is the decomposed Hamiltonian at imaginary time $\tau_n$ which implicitly depends on $\phi_n$. Here $\{\phi_n\}$ represent all possible auxiliary-field configurations on the space-time lattice.

It is advantageous in various ways to map the quantum partition function into a summation of matrix determinants over classical auxiliary-field configurations as shown above. Most importantly, although the determinant in \Eq{determinant} is not guaranteed to be positive definite for all quantum models, the fermion-sign problem can be solved for certain models by choosing appropriate ways of performing Hubbard-Stratonovich transformation. For instance, the attractive-$U$ Hubbard model is a prototype one whose fermion-sign problem can be solved in the framework of DQMC \cite{Hubbard1,Hubbard2,Hubbard3,Hubbard4} by decoupling the Hubbard-$U$ interaction into the density channel. As only the density operators couple with auxiliary fields, the determinant of Boltzmann weight is the product of two equal and real factors, one for spin-up electrons and the other for spin-down electrons, such that it is positive definite. This way of thinking was later generalized to certain quantum models of SU(N) fermions with even $N$ \cite{Assaad-2005}. Although factorization of the Boltzmann weight in DQMC into a product over spin or flavor components can help solving the fermion-sign problem, models which have been identified to be sign-problem-free by this approach are still quite limited. Novel strategies capable of solving fermion-sign problem in more generic classes of models are increasingly desired and, if found, will definitely lead to a great leap forward in deeper understanding key physical properties of those quantum many-body systems.

It is remarkable to witness the enormous progresses achieved in recent years in solving the fermion-sign problem in QMC simulations of fermionic many-body models. In this review, we discuss the recent developments of sign-problem-free QMC methods and their applications in simulations of interacting fermionic models. We shall mainly focus on recent developments in and applications of DQMC algorithms. The rest of this review is organized as follows. In Sec. 2, we review the recent developments of fermion-sign-free QMC simulation in the framework of employing Kramers time-reversal symmetry (TRS). In Sec. 3, we discuss the novel ideas of solving fermion-sign problem by employing the Majorana representation of fermions and the Majorana TRS. We shall also discuss a large class of interesting models whose fermion-sign problems have been solved using the Majorana representation and Majorana TRS. In Sec. 4, we focus on the classification of sign-problem-free symmetry classes in DQMC. In Sec. 5, we briefly review the recent developments of solving fermion-sign problem in other algorithms such as continuous-time QMC \cite{CTQMC2,CTQMC3,CTQMC1,CTQMC4,CTQMC5}. Finally, we end with some concluding remarks in Sec. 6.

\section{Quantum Monte Carlo in complex fermion representation}

\subsection{Kramers time-reversal principle for sign-problem-free QMC}
Even though a general solution of the fermion-sign problem in QMC is NP-hard, many specific models have been successfully identified to be sign-problem-free. As mentioned in the introduction above, appropriate factorization of Boltzmann weights may solve fermion-sign problem in DQMC. Utilizing this approach, many models with interesting features have been investigated by sign-problem-free QMC simulations, including the attractive Hubbard model \cite{Hubbard1,Hubbard2,Hubbard3,Hubbard4,Hubbard5,Hubbard6} and the repulsive Hubbard model on a bipartite lattice at half-filling \cite{honey1,honey2,honey3,honey4}, the Kane-Mele-Hubbard model \cite{KaneMele-1,KaneMele-2,KaneMele-3,KaneMele-4,KaneMele-5,KaneMele-6,KaneMele-7,KaneMele-8, KaneMele-9,KaneMele-10}, and the half-filled Kondo lattice model \cite{Kondo1,Kondo2,Kondo3}.

However, it is certainly not guaranteed that all the Boltzmann weights can be explicitly factorized into two symmetric parts which are real even when the system is at half-filling. For instance, adding a finite Rashba spin-orbit coupling into the attractive Hubbard model would ruin the factorization of the Boltzmann weight into a spin-up part and a spin-down part. Consequently, a more general approach solving fermion-sign problem in QMC was desired. In Ref. \cite{Wu-2005}, Wu and Zhang proposed a more general principle for sign-problem-free QMC simulations. The principle is based on the Kramers TRS of the decomposition of interactions in the Hamiltonian. Note that it is implicitly assumed here that the quantum Monte Carlo is performed in the complex fermion representation; and the particle number after the decomposition is conserved, as shown in $\hat H_0$ and $\hat h(\phi_n)$ above. The principle proposed by Wu and Zhang is captured by the following theorem.

\underline{Theorem}:
For a set of square matrices $H(n)$, if there exists an anti-unitary operator $T$ with $T^2 = -1$ such that
\bea
T^{-1}H(n)T = H(n),
\eea
the eigenvalues of the matrix $B=\mathbb{I}+\prod_n e^{-\Delta\tau H(n)}$ always appear in complex conjugate pairs. Moreover, the determinant of $B$ is positive definite because $\det(B) = \prod_{i} |\lambda_i|^2 >0$.

Here we shall not review the detailed proof of the theorem [see Ref. \cite{Wu-2005} for details].  This theorem may be viewed as the generalized Kramers theorem (as $H_n$ above may be non-Hermitian). We therefore call it Kramers time-reversal principle for sign-problem-free QMC. It is clear that this principle can be applied to more general models to solve their fermion-sign problem compared with the approach of Boltzmann weight factorization. Employing this principle rather than the factorization approach, various models, including the Hubbard models with higher spin and the bilayer Scalapino-Zhang-Hanke model, were shown to be sign-problem-free \cite{Wu-2005,Wu-2004}.

\subsection{Recent applications of Kramers time-reversal principle}
In the past several years, there has been a surge of sign-problem-free QMC studies of strongly correlated systems. QMC simulations which are sign-problem-free using the Kramers time-reversal principle have been recently especially employed to investigate quantum critical fluctuations as well as superconductivity induced by those fluctuations in cuprate or iron-based superconducting materials \cite{Berg-2012}. Antiferromagnetic (AF) order and its fluctuations in metals are largely believed to play an important role in understanding key physics in correlated systems such as cuprates [for a review, see e.g. Refs. \cite{Kivelson-2003,XGWen-2006}] and iron-based superconductors (SCs) [for a review, see e.g. Refs. \cite{Fa-Lee-2011,Mazin-2011}]. It was proposed to use the field theory involving the AF fluctuations and electrons around hot spots to describe low-energy physics of antiferromagnetism in metals near an AF quantum critical point (QCP) \cite{Hertz-1976,Millis-1993,Chubukov-2000,Kotliar-2001,Sachdev-2004}. Nonetheless, due to strong correlations in fermionic gapless degrees of freedom in the metal around the AFQCP \cite{Sachdev-2010,SSLee-2009}, obtaining controlled analytical results in this effective field theory is notoriously challenging. Moreover, the effective field theory suffers from fermion-sign problem which hampers reliable large-scale QMC simulations. Consequently, the solution of this problem remained elusive for a long time, mainly because there were no controlled analytical or numerical approaches available to handle it.

\begin{figure}[t]
\includegraphics[height=1.5in]{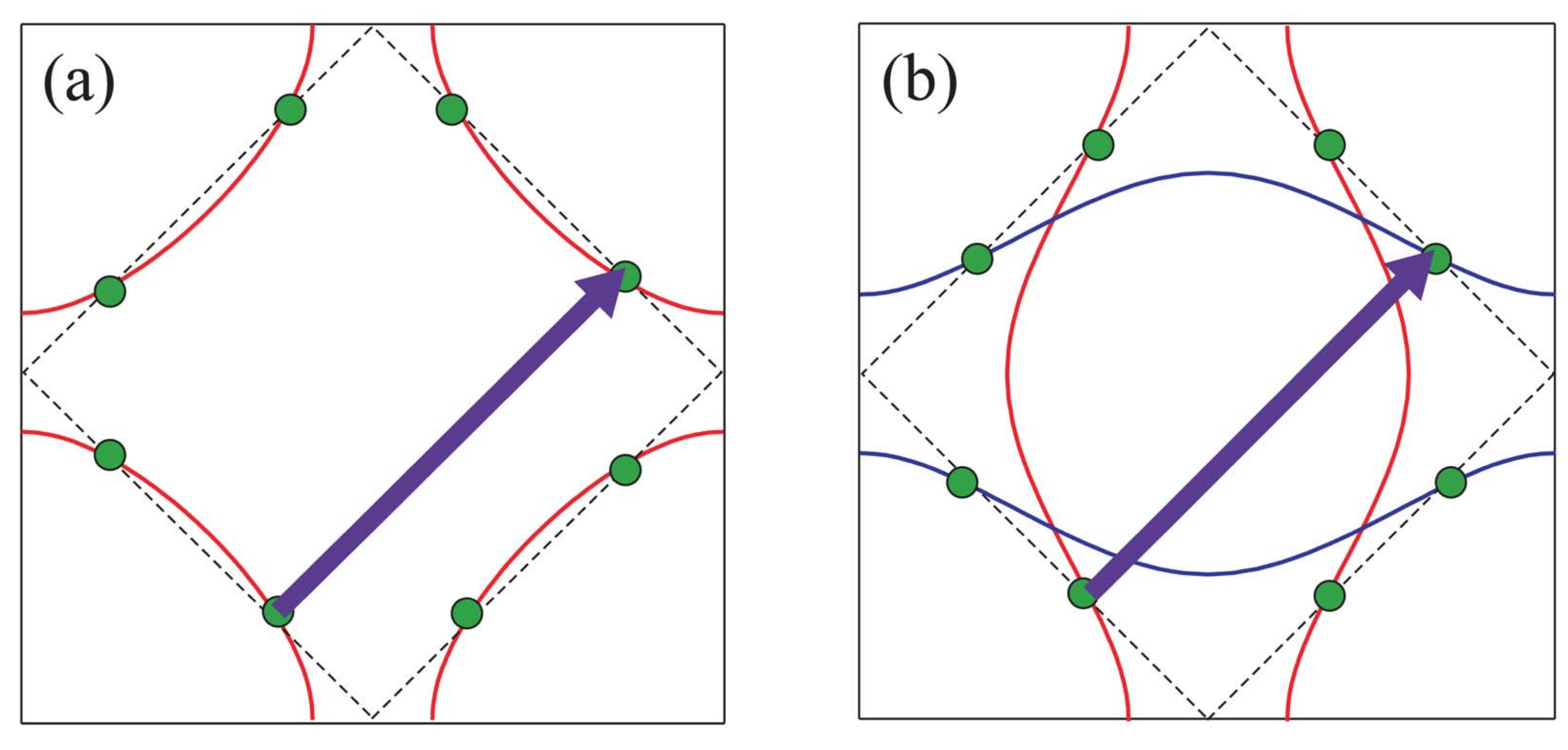}
\caption{(a) A prototypical Fermi surface of the cuprates. The hot spots are denoted by the green dots and the purple arrow indicates the AFM ordering wavevector $(\pi,\pi)$. (b) The Fermi surfaces of the two-band model where the positions of hot spots and the Fermi velocity at hot spots are made to mimic those in the single-band model used to describe the cuprates. }
\label{cuprate}
\end{figure}

Several years ago, Berg, Metlitski, and Sachdev proposed to study critical antiferromagnetic fluctuations in metals by performing sign-problem-free QMC simulations of a different but low-energy equivalent model \cite{Berg-2012}. To study the effects of AF quantum fluctuations in (electron-doped) cuprate which features a single low-energy band of electrons [as shown in \Fig{cuprate}(a)], their strategy was to design a two-band lattice models [as shown in \Fig{cuprate}(b)] such that the AF order-parameter fluctuation bosons scatter from one band to another while the low-energy hot spot physics is identical to the single band model. Utilizing a generalized TRS (in both band and spin spaces), the Boltzmann weight in QMC simulations of the two-band model can be shown to be positive-definite, namely the QMC simulations of the two-band model is sign-problem-free. In converting a single band model to the two-band model, it is important to preserve the low-energy physics near the AF QCP by requiring the same structure of hot-spots connected by AF fluctuations and the same Fermi velocities at those hot spots.
To make it explicit, we write down the action $S= S_{f} + S_{b}+S_{fb} = \int_0^\beta d\tau (L_f + L_b+L_{fb})$ which describes the lattice model with two electron bands designed to mimic the AF phase transition in electron-doped cuprates with one electron band:
\bea{\label{Berg}}
L_f &=& \sum_{\alpha=x,y}\sum_{i,j} \psi^\dagger_{\alpha,i}[(\partial_\tau - \mu)\delta_{ij} - t_{\alpha,ij}]\psi_{\alpha,j},  \nonumber\\
L_b &=& \frac{1}{2} \sum_i \frac{1}{c^2} (\frac{d\vec{\phi}_i}{d\tau})^2 + \frac{1}{2} \sum_{\avg{ij}} (\vec{\phi}_i - \vec{\phi}_j)^2 + \sum_i(\frac{r}{2}\vec{\phi}_i^2 + \frac{u}{4}(\vec{\phi}_i^2)^2), \nonumber\\
L_{fb}&=&\lambda\sum_i \psi^\dagger_{x,i}(\vec{s}\cdot \vec{\phi}_i)\psi_{y,i} + H.c.
\eea
where $\psi^\dag_{\alpha,i}=(\psi^\dag_{\alpha\uparrow,i},\psi^\dag_{\alpha\downarrow,i}$) creates an electron in the $\alpha=x$ or $y$-band on site $i$ with spin polarization $s=\uparrow$ or $\downarrow$, $\vec \phi_i$ denotes the AF fluctuation boson on site $i$, $\tau$ is imaginary time, and $\beta$ is inverse temperature. The parameter $r$ is used to tune the AF phase transition. The band structure using the parameters in Ref. \cite{Berg-2012} is shown in \Fig{cuprate}(b), which preserves the hot-spots\ properties in the original one-band model [as shown in \Fig{cuprate}(a)]. The two-band action above preserves TRS $T= i \sigma^z s^y K$ and $T^2=-1$, where $\vec{s}$ are Pauli matrix acting on spin space, $\vec{\sigma}$ are Pauli matrix acting on band space ($\alpha=x/y$), and $K$ represents the complex conjugation. According to the principle of Kramers TRS, the two-band action in \Eq{Berg} is sign-problem-free. As a consequence, the low-energy physics around the putative AF QCP in the electron-doped cuprate can be simulated by QMC without encountering the notorious fermion-sign problem!

After the pioneering work by Berg, Metlitski, and Sachdev, tremendous efforts \cite{Kivelson-2015,LWYL-2015,Berg-2016,Berg-20162, Ziyang-2015,Fernandes-2016,Ziyang-2016,Kivelson-2016, LWYL-2016, Ashvin-20162, Ziyang-2017, Fernandes-2017} have been made in this direction. In Refs. \cite{LWYL-2015,Berg-2016,Berg-20162}, the phase diagram of \Eq{Berg} and the intertwined orders induced around the AF QCP have been systematically investigated. In Ref. \cite{LWYL-2015}, Li {\it et al.} studied the ground state (or zero-temperature) properties of the AF effective model in \Eq{Berg} by sign-problem-free projector QMC simulations \cite{projector1,projector2,projector3}. The results show that the $d$-wave superconducting pairing is strongly enhanced by AF fluctuations and the ground state exhibits superconducting long-ranged order. Moreover, AF fluctuations can also induce short-range charge density wave (CDW) fluctuation, which was observed in electron-doped cuprates by x-ray experiments \cite{Damascelli-2015,Damascelli-2016, Damascelli-20162}. In Ref. \cite{Berg-2016}, Schattner {\it et al.} studied the same AF effective theory by finite-temperature QMC simulations and obtained consistent results. Remarkably, the phase diagram obtained in AF effective theory is strikingly similar to the phase diagram in some unconventional SC, such as electron-doped cuprate and iron-based SC, strongly suggesting that some essential physics in these materials is captured by the low-energy effective theory featuring AF fluctuations.

Near an AFQCP in metals, the AF fluctuations are strongly coupled with low-energy electrons near Fermi surface. As a consequence, critical properties of such metallic QCP can be dramatically different from the corresponding QCP in insulating systems. Understanding the exotic critical phenomena of the metallic QCP has attracted great interests for many years. Moreover, non-Fermi liquid physics may emerge in the quantum critical regime due to the overdamping of quasi-particles by the fluctuating bosons. QMC simulations, when sign-problem-free, can serve as a controlled approach to address these issues of metallic QCPs. In Ref. \cite{Berg-20162}, Gerlach {\it et al.} performed a sign-problem-free QMC simulation to investigate the critical properties of metallic QCP with fluctuating $O(2)$ spin-density wave (SDW) order parameters. The simulation unveils that the critical exponents of SDW order parameter bosons are remarkably consistent with the Hertz-Millis theory \cite{Hertz-1976,Millis-1993}, such as dynamical exponent $z=2$. The breakdown of Fermi liquid behaviour is also revealed through numerically observing the frequency independent self-energy in their simulations. Recently, Schlief, Lunts, and Lee performed an insightful field-theory study of the O(3) SDW QCP in metals \cite{SSLee-2017,SSLee-2018} employing the velocity ratio as the small parameter and obtained an asymptotically exact fixed point well under theoretical control. The critical exponents of this O(3) SDW metallic QCP are obtained; especially $z=1$ in a certain critical regime. It is desired to perform sign-problem-free QMC simulations in models with a small Fermi velocity ratio to test the analytical results such as $z=1$. Besides SDW quantum critical properties, the quantum critical properties of Ising nematic QCPs in itinerant electrons have been investigated in Refs. \cite{Kivelson-2015, Kivelson-2016} by sign-problem-free QMC simulations. The results reveal exotic quantum critical and non-Fermi liquid behaviours of electrons in the quantum critical regime.

\section{Quantum Monte Carlo in Majorana fermion representation}
\subsection{Majorana time-reversal principle for sign-problem-free QMC}
The Kramers time-reversal principle has achieved many successes in rendering sign-problem-free in complex-fermion quantum Monte Carlo during the past decades. However, the requirements of both Kramers TRS with $T^2=-1$ and particle-number conservation prevent the applications of this principle to more general quantum lattice models. As the minimum dimension of the projective representation satisfying $T^2=-1$ is two, it is thus challenging to apply the principle of Kramers TRS to those models of fermions without internal degree of freedoms such as orbital or spin. For instance, the fermion-sign problem in models of spinless (or spin-polarized) fermions cannot be directly solved in the framework of Kramers TRS. Moreover, the requirement of particle-number conservation makes it difficult to apply this principle to the models which explicitly break charge conservation. In order to resolve these intrinsic difficulties, Li, Jiang, and Yao proposed in Ref. \cite{LJY-2015} to use Majorana-fermion representation to solve fermion-sign problem in QMC simulations. It is the first time that Majorana representation was employed to perform QMC simulations. In Ref. \cite{LJY-2015}, TRS in Majorana representation was utilized to prove the positivity of the Boltzmann weight, which is beyond the framework of Kramers TRS. Additionally, the requirement of particle-number conservation is naturally released in the algorithm using Majorana representation. As a genuinely new QMC algorithm, the Majorana time-reversal principle was employed to solve fermion-sign problem in a large class of interesting models that are beyond the solvability of previous QMC methods. Below, we introduce Majorana QMC and then discuss some applications.

We begin by introducing the basic idea of Majorana QMC. One important motivation of proposing the Majorana QMC was to solve fermion-sign problem in spinless fermionic models \cite{LJY-2015}. It is obvious that no spin degree of freedom in spinless fermion models can be utilized to apply the Kramers time-reversal principle for sign-problem-free QMC. In Ref. \cite{LJY-2015}, by observing that each complex fermion can be represented by two Majorana fermions, Li, Jiang, and Yao proposed to utilize the internal Majorana degrees of freedom to solve fermion-sign problem. Under certain conditions, such as particle-hole symmetry, spinless fermion interactions can be decoupled into the sum of two separated parts each involving only one species of Majorana fermions. When the decomposition further respects TRS in Majorana representation, the traces of two separated Majorana components are complex conjugate to each other such that the Boltzmann is positive-definite, namely sign-problem free.

To explicitly illustrate how the fermion-sign problem can be solved by Majorana QMC, we consider the spinless-fermion model $H = H_0 + H_{I}$  as an example:
\bea
H_0 &=& -t\sum_{\avg{ij}} [ c^\dagger_i c_j + H.c.],  \nonumber\\
H_{I} &=& V\sum_{\avg{ij}} (n_i - \frac{1}{2})(n_j - \frac{1}{2}),
\eea
where $c^\dag_i$ creates a spinless fermion (or spin-polarized electron) on site $i$, $n_i=c^\dag_i c_i$ is the density operator, $t$ denotes nearest-neighbor (NN) hopping amplitude, and $V>0$ labels NN density-density interactive interaction. The Hamiltonian can be defined on any bipartite lattice and the generalization to longer range interactions is straightforward. To solve fermion-sign problem of this model, the Hamiltonian above can be expressed in terms of Majorana fermions with two species because each complex fermion can be rewritten as two Majorana fermions:
\bea
c_{i}=\frac{1}{2}(\gamma_{i}^{1}+i\gamma_{i}^{2}), ~~c^{\dagger}_{i}=\frac{1}{2}(\gamma_{i}^{1}-i\gamma_{i}^{2}),
\eea
where $\gamma^\alpha_i$ ($\alpha=1,2$) are Majorana fermions satisfying $\{\gamma^{\alpha}_i,\gamma^{\tilde \alpha}_{j}\}=2\delta_{\alpha\tilde \alpha}\delta_{ij}$.
After rewriting the Hamiltonian in Majorana representation and performing the standard procedure of Suruki-Trotter decomposition, the interaction $H_{\rm{int}}$ can be decoupled by a HS transformation:
\bea
e^{-\Delta\tau H_{I}}=e^{\frac{V\Delta\tau}{4}(i \gamma_{i}^{1}\gamma_{j}^{1})(i\gamma_{i}^{2}\gamma_{j}^{2})}
= \frac{1}{2}\sum_{\sigma_{ij}=\pm 1}e^{\frac12\lambda \sigma_{ij} (i \gamma_{i}^{1} \gamma_{j}^{1} + i \gamma_{i}^{2}\gamma_{j}^{2})-\frac{V\Delta\tau}{4}}, \label{V1sign}~~~
\eea
where $\Delta\tau$ is the imaginary time slice in the Suruki-Trotter decomposition, $\lambda=\cosh^{-1}[e^{V\Delta\tau/2}]$, and $\sigma_{ij}=\pm 1$ represent auxiliary fields on the NN bond $\avg{ij}$. Note that the HS transformation above is performed in the hopping channel of Majorana fermions, which is a key step for solving the fermion-sign problem. (In conventional complex-fermion QMC, the HS transformation in density channels would encounter fermion-sign problem).

When the system is at half-filling, in terms of Majorana fermions the non-interacting part of the Hamiltonian is given by $H_0=-\frac{t}4 \sum_{\avg{ij}} (i \gamma_{i}^{1} \gamma_{j}^{1} + i \gamma_{i}^{2}\gamma_{j}^{2})$. As a consequence, both free Hamiltonian $H_0$ and the decoupled form of the interacting part are the sum of two separate parts, each of which involves only one species of Majorana fermions. Therefore, the Boltzmann weight $W(\{\sigma\})$ depending on the auxiliary field configuration $\{\sigma\}$ can be written as the product of two factors,
\bea
W(\{\sigma\})&=& W_1(\{\sigma \}) W_2(\{\sigma\}),
\eea
where the factor $W_\alpha(\{\sigma\})$ is obtained by tracing out Majorana component $\alpha$ ($\alpha=1,2$). It is explicitly given by
\bea
W_\alpha(\{\sigma\})&=& \Tr\left[ \prod_{n=1}^{N_\tau} e^{\frac14\widetilde \gamma^{\alpha} h^\alpha (n) \gamma^\alpha} \right] = \pm\left\{\det\bigg[ \mathbb{I}+\prod_{n=1}^{N_\tau} e^{h^\alpha(n)}\bigg]\right\}^{\frac{1}{2}},
\eea
where $\tilde \gamma^\alpha=(\gamma^\alpha)^T=(\gamma^\alpha_1,\cdots,\gamma^\alpha_N)$ and $h^\alpha(n)$ is a $N$$\times$$N$ matrix ($N$ is the number of lattice sites) whose ($ij$)-matrix element is given by
$h^a_{ij}(n) \!=\! i\left[t\Delta\tau +\lambda\sigma_{ij}(n)\right]\delta_{\avg{ij}}$, where $\delta_{\avg{ij}}$ is 1 for $ij$ being NN sites and 0 others. Note that the trace of exponential bilinear Majorana fermions is the square root of a determinant rather that the determinant itself, which is due to the fact that a Majorana fermion carries only half the degree of freedom of a complex fermion.

It is worthy to note that there is sign ambiguity when taking a square root from the above, similar to the case of taking a Pfaffian as a square root of a determinant. Consequently, an additional condition is needed to be sign-problem free. The key observation is that the Majorana Hamiltonian after the HS transformation respects a TRS, which is
defined as $T=GK$ with $K$ being complex conjugation and $G$ being a unitary transformation. Specifically, under the unitary transformation $G$: $\gamma_i^1 \to (-1)^i \gamma_i^2$. Under this Majorana time-reversal transformation, $\widetilde \gamma^1 h^1 (n)\gamma^1 \leftrightarrow \widetilde \gamma^2 h^2(n) \gamma^2$. Consequently, the decomposed quadratic Hamiltonian respects the Majorana TRS $T$. Because of this Majorana TRS, we obtain
\bea
W_{1}(\{\sigma\})= W_2^{*}(\{\sigma\}),
\eea
such that the product of two factors is positive definite: $W(\{\sigma\}) = W_{1}(\{\sigma\}) W_2(\{\sigma\}) = \left|\det\Big[ \mathbb{I}+\prod_{n=1}^{N_\tau} e^{h^1(n)} \Big]\right|>0.$ This proves that fermion-sign problem in such class of models consisting of spinless fermions can be solved by TRS in Majorana representation. Note that the Majorana time-reversal principle for sign-problem-free QMC is quite general, not limited to models of spinless fermions. It is straightforward to generalize this Majorana approach to solve fermion-sign problem in more general models such as systems consisting spin-$\frac12$ electrons \cite{Jiang-2016} and SU($N$) fermions \cite{LJJY-2017}.

\subsection{Applications of Majorana quantum Monte Carlo}
We have shown that fermion-sign problem in the Majorana QMC can be solved by utilizing the Majorana time-reversal principle. Due to the intrinsic properties of Majorana fermions, sign-problem-free models in Majorana QMC include, but are not limited to, two unique classes of models which are normally sign-problematic in complex-fermion QMC. Models in the first class often describe systems of spinless fermions or fermions with odd numbers of species (flavors), whereas models in the second class are interacting models breaking particle-number conservation such as interacting models with explicit pairing terms. Through simulating these novel sign-problem-free models, exotic phases and phenomena have been firmly revealed, including emergent supersymmetry (SUSY) and fermion-induced quantum critical points (FIQCP). In the following, we shall discuss applications of Majorana QMC to sign-problem-free models which exhibit novel physical properties.

Majarana QMC was first applied in Ref. \cite{LJY-2015} to study the CDW quantum phase transition in the spinless fermion model on the honeycomb lattice which features massless Dirac fermions \cite{Wang-2014, LJY-20152, Wessel-2016, Liu-2015, Chandrasekharan-2017, Wang-2016}. Quantum critical behaviours of this quantum phase transition belongs to $N=2$ Gross-Neveu chiral-Ising universality class \cite{Gross-1974}, which were studied using various renormalization group (RG) approaches \cite{RGIsing1,RGIsing2}. However, perturbative calculations of critical exponents in RG analysis may not be that reliable, especially when the number of flavors of two-component Dirac fermions $N$ is small. By performing unbiased sign-problem-free Marajora QMC having large system size, accurate critical exponents have been obtained, which could serve as benchmark for future higher-loop RG calculations. The CDW quantum phase transition of spinless fermions has also been studied on the Lieb lattice via Majorana QMC in Ref. \cite{Lang-2017}, where the particle-hole symmetry is spontaneously broken at low temperature by developing a finite CDW order. It is interesting to observe that the spontaneous breaking of particle-hole symmetry can lead to a ground-state particle number deviated from half-filling, which may shed light on simulating finite-doping system without fermion-sign problem.

Majorana QMC was also applied to investigate interacting models of SU($N$) models, especially those with odd $N$. For instance, Majorana QMC simulations of interacting models of SU($N$) fermions on the honeycomb lattice at half-filling revealed an exotic QCP dubbed as FIQCP in Ref. \cite{LJJY-2017}. The quantum phase transition is characterized by a Kekule valence-bond-solid (VBS) order, which is putatively a first-order transition according to the Landau cubic criterion that continuous phase transition is prohibited when cubic term of order parameters is allowed by symmetry in Landau-Ginzburg (LG) free energy.   However, the non-trial coupling between the $Z_3$ order parameter and massless Dirac fermions may change characters of the phase transition dramatically. In Ref. \cite{LJJY-2017}, it was argued by RG analysis that when the number of flavors of Dirac fermion $N$ is sufficiently large, the low-energy effective theory describing the Kekule-VBS phase transition flows to a fixed point where the cubic term is renormalized to zero, indicating that the phase transition is continuous, rather than first-order. This quantum critical point is called fermion-induced quantum critical point \cite{LJJY-2017,Jian-2017,Jian-2017b,Herbut-2016,Congjun-2016}. To verify the scenario of FIQCP obtained by perturbative RG analysis, Li {\it et al.} constructed a microscopic model of $SU(N)$ fermions on the honeycomb lattice, which features a phase transition between Dirac semimetal and Kekule VBS phase and which is sign-problem-free in the framework of Majorana QMC. The state-of-the-art QMC simulations show convincing evidences of FIQCP for $N=2,3,\cdots,6$. The FIQCP verified by sign-problem-free Majorana QMC paves a new route to realize exotic quantum phase transitions beyond conventional LG paradigm.

In Majorana-based QMC, fermion-sign problem can be solved for certain models without particle-number conservation. It is therefore natural to apply Majorana QMC algorithm to investigate interacting effects in superconductors, namely Bogoliubov-de Gennes (BdG) Hamiltonians with interactions. In Ref. \cite{LJY-20162}, sign-problem-free Majorana QMC was applied to study whether spacetime supersymmetry emerges at the edge quantum critical points induced by interactions in a 2D time-reversal invariant (TRI) topological superconductor. In Ref. \cite{Ashvin-2014}, it was argued by Grover, Sheng, and Vishwanath from RG analysis that 1+1D spacetime SUSY emerges at the edge QCP of spontaneous TRS breaking on the edge of 2D topological superconductors. To test possible existence of an edge QCP and associated emergent SUSY in a 2D lattice model, we introduce the following minimal model of interacting TRS topological SC in Ref. \cite{LJY-20162}:
\bea\label{tsc}
H \!=\! \sum_{ij}\sum_{\sigma=\uparrow,\downarrow} \!\big[\!-t_{ij} c^\dagger_{i\sigma} c_{j\sigma} \!+\!\Delta_{ij,\sigma}c^\dagger_{i\sigma}c_{j\sigma}^\dagger \!+\! h.c.\!\big]\! \!-\!U\sum_{i} n_{i\uparrow} n_{i\downarrow},~~~
\eea
where the noninteracting part above represents a BdG description of 2D TRI topological superconductors with $\Delta_\uparrow$ ($\Delta_\downarrow$) denoting $p+ip$ ($p-ip$) pairing between two spin-up (spin-down) electrons. Importantly, after an appropriate HS transformation the decoupled Hamiltonian satisfies the Majorana time-reversal principle such that Majorana QMC simulations are free from the fermion-sign problem.
As is revealed by large-scaled QMC simulations in Ref. \cite{LJY-20162}, when the interaction $U$ increases spontaneous TRS breaking occurs on the edge at a critical value $U^{\rm{edge}}_c$ while the system's bulk remains gapped. Moreover, various evidences of emergent spacetime SUSY at this edge QCP were obtained. For instance, several critical exponents obtained from QMC are consistent with the exact values associated with the putative $\mathcal{N}=1$ SUSY in 1+1D. More importantly, sufficiently close to the edge QCP, the masses of fermion and boson are found to be equal to each other within error-bar, which is a hallmark of emergent SUSY. It is the first intrinsically-unbiased numerical study of a 2D lattice model revealing evidences of emergent SUSY at its edge QCP. More recently, sign-problem-free QMC simulations in Majorana representation in Ref. \cite{LY-20171} obtained convincing evidences of emergent ${\mathcal N}$=2 2+1D SUSY, proposed theoretically in Refs. \cite{Balents-1998,Ashvin-2014,SSLee-2014}. It would be interesting to study emergent 3+1D SUSY in quantum many-body models \cite{Jian-2015} by sign-problem-free QMC in the future.

In addition to examples discussed above, a number of other interesting models have been investigated through sign-problem-free Majorana QMC. For instance, Majorana QMC was employed to study the ground-state stability of charge-4e superconductivity \cite{Berg-2009,Berg-20092,Jiang-2016} and the edge stability of 2D topological insulators under strong interactions \cite{LY-20172}. More recently, sign-problem-free models describing fermions coupled to fluctuating $Z_2$ gauge fields are constructed to investigate exotic deconfined phases and phase transitions in fermionic systems \cite{Ashvin-2016,Assaad-2016}, where for odd number of fermion species the proof of the absence of fermion-sign problem relies on the Majorana representation introduced in Ref. \cite{LJY-2015}. Fermionic models which are sign-problem-free in Majorana representation may serve as a platform to study certain entanglement properties of interacting fermionic system \cite{Trebst-2016} as well as detect quantum phase transitions by machine learning \cite{Trebst-2017}.

\section{Classification of sign-problem-free symmetry classes}
\subsection{The classification scheme}
We have shown that fermion-sign problem may be solved by using TRS of the HS transformation in complex fermion representation as well as in Majorana representation. In order to have a deeper understanding of fermion-sign problem as well as design more general sign-problem-free models exhibiting interesting physics, it is desirable to find a fundamental guiding principle for sign-problem-free algorithms. In Ref. \cite{LJY-2016}, Li, Jiang, and Yao proposed to classify sign-free and sign-problematic models in QMC according the the set of anti-unitary (loosely called ``time-reversal'' hereafter) symmetries in Majorana representation that all the decomposed Hamiltonians $H_n$ respect. The idea of employing anti-commuting TR (or anti-unitary) symmetries in Majorana representation to classify fermion-sign problem of random Majorana bilinear operators is, in spirit, similar to the one used by Kitaev in obtaining a ``periodic table'' classification of topological insulators and topological superconductors \cite{Kitaev-2009}. Since Majorana fermion operators are real, a Majorana-time-reversal symmetry can be represented by $T^\pm=U^\pm K$, where $(T^\pm)^2=\pm 1$ and $U^\pm$ is a real orthogonal matrix satisfying $(U^\pm)^T=\pm U^\pm$. The decomposed Hamiltonian $\hat h_{\tau_n} = \gamma^T h_n \gamma$ expressed in Majorana representation may be classified systematically according to the maximal set of anti-commuting Majorana TRS ${\mathcal C}=\{T_1^{p_1},\cdots,T_m^{p_m}\}$ they respect, namely $[T_j^{p_j},h_n]=0$ and $T_i^{p_i}T_j^{p_j}+T_j^{p_j}T_i^{p_i}=p_i2\delta_{ij} $, where $m$ is the number of TRS in a symmetry class and $p_i=\pm$. Because of possible sign choices of $p_i=\pm$, there are totally $m+1$ distinct symmetry classes for each $m$. In the following, we shall show whether a given symmetry class is sign-problem-free or sign-problematic.

\subsection{Two fundamental sign-problem-free symmetry classes: Majorana class and Kramers class}
In Ref. \cite{LJY-2016}, it was found that there are two fundamental sign-problem-free symmetry classes and all other sign-problem-free symmetry classes must respect higher symmetries. We shall sketch the proof below. Basically, we check all symmetry classes starting with $m=0$. Here, $m=0$ means that the Hamiltonian $h_n$ does not respect any Majorana TRS, which is obviously sign-problematic. For $m=1$, there are two distinct symmetry classes: $\{T_1^+\}$ and $\{T_1^-\}$. A single Majorana TRS is not enough to guarantee that they are sign-problem-free due to the possible minus sign appearing in taking the square root of a determinant by tracing out Majorana fermions. For instance, $\Tr \left[\exp(x\gamma^1\gamma^2)\right] =2\cos x$, which is negative for certain range of $x$, for example $x\in(\frac{\pi}{2},\pi)$, even though the Majorana-bilinear operator $x \gamma^1 \gamma^2$ respects the $T^-_1$ symmetry defined as: $\gamma^1$$\to$$\gamma^2$, $\gamma^2$$\to$$-\gamma^1$, plus complex conjugation.

Consequently, to be possibly sign-problem-free, we need to consider symmetry classes with $m\geq2$. For $m=2$ there are three distinct symmetry classes: $\{T_1^+,T_2^+\}$, $\{T_1^+,T_2^-\}$, and $\{T_1^-,T_2^-\}$. The symmetry class $\{T_1^+,T_2^+\}$ may encounter the sign problem. In the following, we show that the two symmetry classes $\{T_1^+,T_2^-\}$ and $\{T_1^-,T_2^-\}$ are sign-problem-free. The former is dubbed as ``Majorana class'' while the latter ``Kramers class''.

\textbf{Majorana class}: If all $h_n$ respect two Majorana TRS $T_1^+=U_1^+K$ and $T_2^-=U_2^-K$, the Boltzmann weight $W=\Tr\prod_{n=1}^{N_\tau} \exp[\hat{h}_n]$, where $\hat h_n=\gamma^T h_n\gamma$, must be positive-definite.

Proof: From these two time-reversal symmetries, one can construct a unitary symmetry $P=T_1^+T_2^-=U_1^+U_2^-$.  As $[P,h_n]=0$, we can use $P$ to block-diagonalize all $h_n$. Namely, one can block-diagonalize all $h_n$ into the following form:
\bea\label{block}
 h_n \to h'_n=\left(\begin{array}{cc}
B_n & 0 \\ 0 & B_n^*
\end{array}\right).
\eea
Consequently, the Boltzmann weight $W=\Tr\prod_{n=1}^{N_\tau} \exp[\hat{h}_n]$ are positive definite, as required by the complex conjugation between the two decoupled blocks $B_n$ and $B_n^\ast$, as shown in Ref. \cite{LJY-2015}. As charge-conservation is not required in this symmetry class, it is a new {\it sign-problem-free} symmetry-class dubbed as ``Majorana class''.

\begin{table}[b]
\centering
\caption{The classification of fermion-sign problem for symmetry classes defined by the set of anti-commuting Majorana TR symmetries $\{T_1^{p_1},T_2^{p_2},\cdots,T_n^{p_n}\}$ they respect. }\label{table}
\begin{tabular}{|c|c|c|}
\hline
\tabincell{c}{Number of anti-commuting
\\ Majorana time-reversal symmetries}&\bfseries Sign-problem-free& Sign-problematic \\
\hline
$0$ & none & \tabincell{c}{no Majorana \\time-reversal symmetry}  \\
\hline
$1$ & none & \tabincell{c}{$\{T^+\}$, $\{T^-\}$} \\
\hline
$2$ & \tabincell{c}{$~\{T^+_1,T^-_2\}$=Majorana class\\$\{T^-_1,T^-_2\}$=Kramers class}&$\{T^+_1, T^+_2\}$ \\
\hline
$3$ & \tabincell{c}{$\{T^+_1,T^+_2,T^-_3\}$\\$\{T^+_1,T^-_2,T^-_3\}$\\$\{T^-_1,T^-_2,T^-_3\}$} & $\{T^+_1,T^+_2,T^+_3\}$  \\
\hline
$4$ or more & all  & none  \\
\hline
\end{tabular}
\end{table}

\textbf{Kramers class}: If all $h_n$ respect two Majorana TRS $T_1^-=U_1^-K$ and $T_2^-=U_2^-K$, the Boltzmann weight $W=\Tr\prod_{n=1}^{N_\tau} \exp[\hat{h}_n]$ must be positive-definite.

Proof: From $T_1^-$ and $T_2^-$ symmetries, a unitary symmetry $Q=T_1^-T_2^-$ can be derived. Because $Q$ is anti-symmetric, namely $Q^T=-Q$, one can construct a charge operator $\hat Q=\gamma^T (iQ) \gamma$ such that $[\hat Q,\hat h_n]=0$. It is clear that the combination of time-reversal $T_1^-$ and charge-conservation $Q$ in this symmetry class is equivalent to the Kramers-symmetry since $[T_1^-, iQ]=0$ and $(T_1^-)^2=-1$. The absence of fermion-sign-problem has been shown according to the observation of the complex conjugate pairs in eigenvalues in Ref. \cite{Wu-2005}. Because of the Kramers TRS, this symmetry class is dubbed as ``Kramers class''.

For symmetry classes $\{T_1^+,T_2^+\}$ and $\{T_1^+,T_2^+,T_3^+\}$, it's easy to find sign-problematic examples. The detailed discussions are given in Ref. \cite{LJY-2016}. For symmetry classes with $n\ge 3$ except $\{T_1^+,T_2^+,T_3^+\}$, it turns out that all of them contain the symmetries in either Majorana class or Kramers class as a subset. Consequently, all of them are sign-problem-free. For instance, the symmetry class $\{T_1^+,T_2^+,T_3^-\}$ contains $\{T_1^+,T_3^-\}$ as a subset, namely it has higher symmetry than the Majorana class and is then sign-problem free.

Combining all the results above, it has have shown that there are two and only two \textit{fundamental} sign-problem-free symmetry classes: Majorana class and Kramers class.  The ``table'' of symmetry classes which are sign-problem-free or sign-problematic in QMC simulations is summarized in Table \ref{table}.

\subsection{Sign-problem-free conditions beyond symmetry principle}
By systematically classifying fully random matrices according to the set of anti-commuting Majorana TR symmetries, it has been rigorously shown that there are two and only two fundamental sign-problem-free symmetry classes. However, it is possible that certain lower symmetry classes with special non-symmetry conditions may be sign-problem-free. For instance, in Ref. \cite{Xiang-2016}, Majorana reflection positivity \cite{reflection1,reflection2} condition was insightfully added to the symmetry class $\{T^+\}$ to solve fermion-sign problem. Reflection positivity was initially introduced in quantum field theory and has many applications in quantum many-body problems. Employing the method of spin reflection positivity, Lieb proved the uniqueness of the ground state of Hubbard model at half-filling \cite{Lieb-1989}. The requirement of reflection positivity puts strong constraints on the sign structure of ground state wave function \cite{Xiang-2015} which is intimately related to fermion-sign problem in QMC. It is therefore natural to connect reflection positivity to fermion-sign problems in QMC simulations.

Firstly, we briefly discuss Majorana reflection positivity. The operator $\hat{O}$ is reflection positive if it satisfies the condition: $\Tr[\hat{Q} \theta(\hat{Q}) \hat{O}] \geq 0$ for any operator $\hat{Q}$ spanned in Majorana spaces, where $\theta$ is Majorana reflection operation. Obviously, the positivity of this quantity is useful in solving fermion-sign problem in QMC simulation. For the generic form of Boltzmann weight in DQMC: $W = \Tr \left[\prod_{n=1}^{N_\tau} e^{\Delta\tau \hat{H}(n)}\right]$, where $\hat{H}(n) = \gamma^T H (n) \gamma$ is the bilinear Majorana operator after HS decomposition, it was shown in Ref. \cite{Xiang-2016} that $W$ is positive-definite if all the matrices $H(n)$ satisfy the following Majorana reflection positivity condition:
\bea\label{V}
H(n) = \left(\begin{array}{cc}
A_n & i B_n \\ -i B^T_n & A^*_n
\end{array}\right),
\eea
where $A_n$ and $B_n$ are $N\times N$ matrices with $A_n$ being anti-symmetric and $B_n$ being positive semi-definite. See Ref. \cite{Xiang-2016} for the details of the proof. Note that the requirement of $B_n$ being positive semi-definite cannot be guaranteed by any symmetries. In other words, this reflection-positivity criterion of being sign-problem-free in Majorana QMC goes beyond the symmetry principle of solving fermion-sign problem discussed in previous sections.

The Hamiltonian $\hat{H}(n) = {\gamma}^T H(n) {\gamma}$ in \Eq{V} respects a Majorana TRS $T^+ = \sigma^x \otimes I_N $, and thus belongs to sign-problematic symmetry class $\{T^+\}$ without requiring that $B_n$ is positive semi-definite. It is remarkable that the positive semi-definite condition of $B_n$ would guarantee sign-problem-free in the symmetry class $\{T^+\}$. Note that such conclusion does not contradict the symmetry classification obtained by Majorana TRS since the positive semi-definite condition of $B_n$ is not a symmetry requirement. Nonetheless, it inspires one to look for new types of sign-problem-free models which are beyond the symmetry principles. Indeed, recently certain mathematical technique beyond the symmetry principles are introduced to solve fermion-sign problems such as semigroup approach \cite{Wei-2017}. We believe that studying sign-problem-free conditions beyond symmetry principles is one promising direction for solving fermion-sign problem.

\section{Recent developments in other quantum Monte Carlo algorithms}
The algorithm of DQMC utilizes discretization of imaginary time by Suruki-Trotter decomposition, which inevitably introduces a Trotter error which is order of $(\Delta\tau)^2$ and can be made sufficiently small by decreasing $\Delta\tau$. About two decades ago, a novel algorithm distinct from Trotter-discretizing imaginary time was proposed. This new algorithm directly samples the diagrammatic expansions and was therefore called as diagrammatic QMC \cite{CTQMC1,CTQMC2,CTQMC3,CTQMC4,CTQMC5,CTQMC6,CTQMC7}. To distinguish it from other QMC algorithms based on discretizing imaginary time, diagrammatic QMC is also referred as continuous-time QMC (CTQMC). For different quantum many-body systems, various approaches of diagrammatic expansion have been developed. For instance, Sandvik \cite{Sandvik-2003} developed a simple and efficient algorithm of stochastic series expansion (SSE) QMC, based on high-temperature expansion, to explore interacting spin systems. In fermionic systems, one standard approach of diagrammatic expansion is diagrammatic determinant QMC(DDQMC) \cite{CTQMC5}, the basic procedure of which is to perform time-dependent series expansion of the interacting part of a Hamiltonian. Take spinless fermionic model with NN density-density interaction for example:
\bea{\label{CTQMC}}
Z = \Tr[e^{-\beta(H_0 + H_{I})}] = Z_0\sum_n \frac{(-1)^n}{n!}\avg{S_I^n}_0,
\eea
where $S_I =\int_0^\beta d\tau H_{I}= V \int_0^\beta d\tau \sum_{\avg{ij}} (n_{i,\tau} - \frac{1}{2})(n_{j,\tau} - \frac{1}{2})$, $H_0$ is non-interacting Hamiltonian of spinless fermionic model, $\langle {\hat{O}}\rangle_0 = \mathcal{T} \Tr[ e^{-\beta H_0} \hat{O}]$, and $Z_0 = \Tr [e^{\beta H_0}]$. Employing Wick's theorem, one can express the expectation values in terms of the determinant of non-interacting Green's function matrix. In general, the Boltzmann weight $\frac{(-1)^n}{n!}\avg{S_I^n}_0$ is not positive definite for a generic $S_I$ such that fermion-sign problem will appear in QMC simulations. Similar to DQMC, fermion-sign problem may be avoided in CTQMC for specific models such as Anderson model at half-filling by utilizing spin degree of freedoms. Recently, using the novel idea of fermion bag \cite{Chandrasekharan-2010,Chandrasekharan-2012, Chandrasekharan-2014}, fermion-sign problem in the spinless fermionic model at half-filling can be solved by the observation that the sign $(-1)^n$ is compensated by the sign of determinant in \Eq{CTQMC} when interactions satisfy some specific conditions. Moreover, interesting models such as $SU(3)$-symmetric attractive Hubbard models on bipartite lattices can be proven to be sign-problem-free utilizing the algorithm of fermion bag. The fermion bag algorithm was recently employed to explore the CDW quantum phase transition in spinless Dirac fermion systems \cite{Wang-2014,Wessel-2016,Chandrasekharan-2017,Wang-2016} and study the phase diagram of mass-imbalanced Hubbard model \cite{Liu-2015}.

It turns out that fermion-sign problem in certain class of models, including spineless fermionic $t$-$V$ model, can be solved in both DQMC by Majorana algorithm and in CTQMC by fermion bag algorithm. For such class of models, one may ask if the solutions of fermion-sign problem in DQMC and in CTQMC can be understood in a unified way. Indeed, in Ref. \cite{Wang-2015}, Wang {\it et al.} made an interesting observation that special properties of split orthogonal groups could help unify these two approaches of sign-problem-free QMC for this class of models, although these two approaches are qualitatively different when considering more general models. The split orthogonal group $O(N,N)$ is formed by $2N$$\times$$2N$ real matrices which preserve the matric $\eta = \textrm{diag}(1,\dots,1,-1,\dots,-1)$: $ M^T \eta M = \eta$. It was shown in Ref. \cite{Wang-2015} that for a series of random real matrices $A_n$ satisfying $\eta A_n \eta = - A_n^T$ [namely $e^{A_n}\in O(N,N)$],
\bea
\det(\mathbb{I}+ \prod_n e^{A_n}) \geq 0.
\eea
In both DQMC and CTQMC, if the decoupled Hamiltonians involving auxiliary fields satisfy the split orthogonal group condition above, the QMC simulations are sign-problem-free. In this regard, split orthogonal group could help unifying solutions of fermion-sign problems in DQMC and CTQMC when the decomposed Hamiltonian respects particle-number conservations. In other words, models which are in the sign-problem-free Majorana class and which are sign-problem-free using Majorana reflection positivity can go beyond the split orthogonal group algorithm as they do not require particle-number conservation.

Besides efforts towards solving fermion-sign problem in broader classes of models, many important technical progresses of QMC algorithms have been made in recent years. For instance, Iazzi {\it et al.} developed a highly efficient CTQMC algorithm \cite{Troyer-LCTQMC1}, the computational time of which scales linearly with inverse temperature. In Ref. \cite{Troyer-LCTQMC2}, Wang {\it et al.} develop the corresponding zero-temperature version of the algorithm. Efficient approaches of computing Renyi entanglemental entropy and fidelity susceptibility, which are useful to identify quantum phases and phase transitions, have been developed based on CTQMC algorithms \cite{Wang-fidelity,Wang-entropy}.

\section{Outlook}
The fermion-sign problem is one of the most important problems in theoretical physics, any (partial) solution of which could lead to a great leap forward in understanding strongly correlated systems. We have concisely reviewed recent progresses in solving fermion-sign problem in fermionic QMC as well as their applications to interesting fermionic models. Although the recent developments of solving fermion-sign problem can only be applied to models with certain conditions, we believe that novel ideas involved in these approaches could open up more inspiring directions in the  future. For instance, TRSs in Majorana representations have been introduced to solve fermion-sign problem. We believe that generalizing the time-reversal principle in Majorana representation to more general symmetries or to other novel representation is a challenging and interesting direction to pursue. Furthermore, utilizing the combinations of symmetry principles and novel mathematical techniques, such as reflection positivity, to solve fermion-sign problem in QMC simulations is another promising direction. More recently, intrinsic connections between fermion-sign problem and gravitational anomaly have been discussed in an insightful work \cite{Ringle-2017}. We believe that establishing deep connections between fermion-sign problems and other aspects of physics is also a subject of great importance. Understanding of fermion-sign problem in QMC simulations is far from complete, both in the level of mathematics and physics. We expect that more powerful mathematical tools and novel physical insights will be introduced to help us understand fermion-sign problem as well as physics of interacting quantum many-body systems.

\section*{ACKNOWLEDGMENTS}
We would like to thank Xun Cai, Shao-Kai Jian, Steve Kivelson, Dung-Hai Lee, Chris Mendl, Abolhassan Vaezi, Fa Wang, Shi-Xin Zhang, and especially Yi-Fan Jiang for collaborations related to this review, as well as to Fakher Assaad, Erez Berg, Alexei Kitaev, Zi-Yang Meng, Matthias Troyer, Lei Wang, Zheng-Yu Weng, Congjun Wu, Tao Xiang, and Shou-Cheng Zhang for helpful discussions about related works. We also acknowledge support from the National Natural Science Foundation of China under Grant Number 11474175 (Z.-X.L. and H.Y.), the Ministry of Science and
Technology of China under Grant Number 2016YFA0301001 (H.Y.), and the National Thousand-Young-Talents Program (H.Y.).

\end{document}